\documentclass[twocolumn,prb]{revtex4}

\usepackage{graphicx}
\usepackage{dcolumn}
\usepackage{bm}

\begin{document}

\title{Interlayer exchange coupling and giant magnetoresistance in Fe/V (001) superlattices}

\author{A. Broddefalk, R. Mathieu and P. Nordblad}
\affiliation{Department of Materials Science, Uppsala University, Box
534, SE-751 21, Uppsala, Sweden }

\author{P. Blomqvist and R. W\"appling}
\affiliation{Department of Physics, Uppsala University, Box
530, SE-751 21, Uppsala, Sweden}

\author{J. Lu and E. Olsson}
\affiliation{Department of Materials Science, Uppsala University, Box 534, SE-751 21, Uppsala, Sweden}

\date{\today}

\begin{abstract}
Magnetization and magnetoresistivity studies of Fe/V (001) superlattices are 
reported. The first giant magnetoresistance peak with 
respect to the vanadium and iron layer thicknesses is investigated. The interlayer antiferromagnetic 
coupling strength is found to show a peak at a vanadium layer thickness of 13 atomic monolayers
($\approx$ 20 \AA) with a full width at half maximum of about 2 monolayers. The antiferromagnetic coupling shows a maximum at an iron layer 
thickness of about 6 monolayers ($\approx$ 9 \AA) for series of superlattices with vanadium thicknesses
around 13 monolayers. The magnitude of the giant magnetoresistance shows a similar 
variation as the antiferromagnetic coupling strength.   
\end{abstract}

\maketitle

\section{INTRODUCTION}
The interlayer exchange coupling (IEC) of ferromagnetic layers through a non-magnetic metal has attracted a lot of attention in the last decade, in connection to the giant magnetoresistive (GMR) effect observed in antiferromagnetically coupled layers\cite{gmr,gmr2}. This interaction has been shown to oscillate
between ferromagnetic (FM) and antiferromagnetic (AF) coupling when varying
the spacer layer
thickness\cite{osc}. The IEC has
also, both experimentally and theoretically,\cite{iec} been shown  to
depend on the thickness of the magnetic
layers. In this case, the coupling coefficient does not necessarily 
change sign but may only show a varying magnitude with increasing layer thickness. Similar variations of the GMR ratio with the magnetic layer
thicknesses have been reported.\cite{gmr}
Fe/V (iron/vanadium) superlattices have been shown to couple antiferromagnetically
for Fe(3 monolayers)/V(12-16 monolayers),\cite{gbergaf} and in a series of Fe (10
{\AA})/V($t_{\rm V}$), oscillations were found with a
maximum in the strength of the antiferromagnetic coupling at V layer 
thicknesses of
$t_{\rm V}$ = 22, 32 and 42 {\AA}.\cite{schwickert} One atomic monolayer (ML) of a Fe/V (001) 
superlattice amounts to about 1.5 {\AA}.

In this paper, the influence of the thickness of the Fe layers
on the IEC and the GMR of Fe/V (001) superlattices near the first AF coupling peak (V$\approx$13 ML) are examined.

\section{EXPERIMENTAL}
\subsection{Sample preparation and characterization}
The Fe/V superlattices (SL) were grown in a three target magnetron sputtering system with a base pressure of 10$^{-10}$ torr. The polished MgO (001) substrates (10x10x0.5 mm$^3$) were ultrasonically precleaned in ethanol, isopropanol and acetone before they were outgassed at 800 $^o$C for 30 minutes. The sputtering gas was Ar with a purity of 99.9999\% and the targets used were Fe(99.95\%), V(99.7\%) and Pd(99.95\%). The sputtering gas pressure was 5.0 mtorr and the substrate holder temperature was 400 $^o$C. The sample holder was electrically isolated from ground potential and rotated ($\sim$ 30 rpm) during deposition to prevent thickness gradients. The epitaxial relationship between Fe and MgO (001) is Fe[001] $\parallel$ MgO [001] and  Fe[110] $\parallel$ MgO [010]. This arrangement gives a nominal lattice mismatch of 3.5\%. On the substrate, Fe and V were alternately deposited by using computer-activated shutters. The layer thicknesses were monitored by the deposition time. Typical deposition rates of Fe and V were 0.65 \AA/s and 0.45 \AA/s respectively. The samples were capped with palladium (Pd) to avoid oxidation. In this paper we use the nomenclature Fe($X$ ML)/V($Y$ ML) , where $X$ and $Y$ indicate the nominal thicknesses of the Fe and V layers in atomic monolayers, respectively. The Fe thickness ranged from 3 to 13 ML ($X$ = 3, 5, 6, 9 and 13 ML), while V thicknesses ($Y$ = 11, 12, 13, 14 and 15 ML) were chosen around the first AF coupling peak near V$\approx$13 ML\cite{gbergaf}.\\
The structural quality of the SL were investigated by XRD. The measurements were carried out in the low-angle region ($2\theta$=1-20$^o$) and in the high-angle region ($2\theta$=20-100$^o$). A Siemens D5000 powder diffractometer was used with the beam defined by 0.3$^o$ divergence and receiving slits. For full width at half maximum (FWHM) measurements, the beam was defined by 0.05$^o$ slits. The Cu K$_\alpha$ radiation was monochromatized by a secondary graphite monochromator. The SUPREX model was used in order to determine the Fe/V interface quality\cite{refsuprex}.\\
Table~\ref{table1} gives for all SL the number of atomic monolayers and repetitions, the thickness of Pd, as well as the nominal and measured superlattice periods $\Lambda$. The nominal $\Lambda$ is estimated from $\Lambda$ = $X \times$a$_{Fe}$/2+$Y \times$a$_V$/2, where a$_{Fe}$=2.8664 \AA and a$_V$=3.0274 \AA are the lattice parameters of Fe and V respectively. The measured $\Lambda$ value is obtained from the XRD measurements. The error on the superlattice period $|$$\Lambda$(nominal) - $\Lambda$(measured)$|$ is small, and amounts on average to 0.35 \AA, i.e. less than 0.25 ML.\\
One of the superlattices, Fe(9 ML)/V(13 ML) was characterized by transmission electron microscopy
 (TEM). A cross-section specimen was prepared by gluing two thin film sample pieces face to face and subsequently cutting slices from the sandwich. Each slice was mechanically ground on both sides to a thickness of 100 $\mu$m. The slice was then dimpled to a thickness of 10 $\mu$m at the specimen center whereafter the specimen was ion milled until electron transparency. The TEM analysis was carried out using both a Tecnai F30 ST field emission gun TEM operated at 300 kV with a Gatan Imaging Filter and a Jeol 2000 FXII TEM operated at 200 kV.

\subsection{Magnetization and magnetoresistance measurements}
Hysteresis loops were recorded for all SL at 10 K in a Quantum Design MPMS5 Superconducting QUantum Interference Device (SQUID) magnetometer. The magnetic field was applied along the [100] and
[110] directions of the Fe layers. The absolute value of the magnetization has been
calculated using the total volume of the Fe layers in each SL as the magnetic
volume, neglecting any influence from induced moments in the V layers.\cite{schwickert} 
For the SL where the coupling was
found to be ferromagnetic and an in-plane anisotropy was observed,
the anisotropy constant $K$ proportional to the energy difference between the [110]
and the [100] direction was deduced from the enclosed area of the two
magnetization curves in the first quadrant of the magnetization vs.
applied field curves. Since the field was applied in-plane, where the
shape anisotropy is small, no correction of the field for
demagnetization effects was done. For the SL where
antiferromagnetic coupling was observed, the coupling strength was
estimated from
$J=\mu_0M_sH_{\rm sat}t_{Fe}/4$, where $t_{Fe}$ is the thickness of the Fe
layers and $H_{\rm sat}$ is the saturation field.\cite{gbergaf}
Resistivity $\rho$($H$, $T$, $\theta$) and magnetoresistance (MR) were
measured using a standard four-probe method and a Maglab 2000 system
from Oxford Instruments with a rotationary probe. The
magnetoresistance was recorded in the current-in-plane (CIP)
geometry. $\theta$ refers to
the angle between the current and the in-plane magnetic field. The
resistance was deduced for H $\Vert$ I ($\theta$=0) and H $\perp$ I
($\theta$=90$^o$) by rotating the sample and always feeding the current
between the same contacts. The MR is defined as $\Delta \rho/\rho_0$ = ($\rho_0-\rho_{sat}$)/$\rho_0$, with $\rho_0$=$\rho$($H$=0) and $\rho_{sat}$=$\rho$($H$=$H_{sat}$).

\section{RESULTS AND DISCUSSION}
\begin{figure}
\includegraphics[scale=0.50]{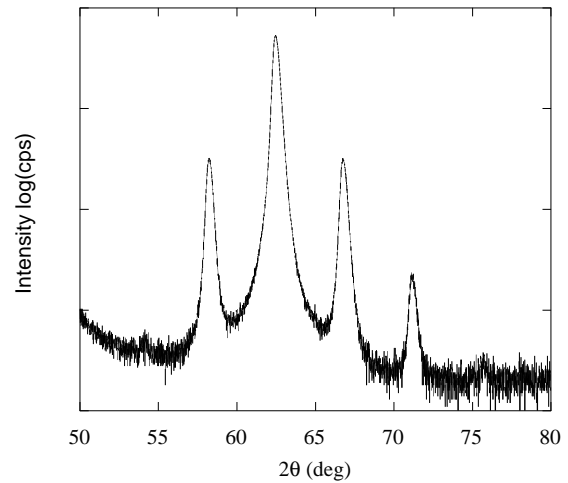}
\caption{High-angle x-ray diffraction scan from the Fe(3 ML)/V(13 ML) superlattice.}
\label{figxray1}
\end{figure}

\begin{figure}
\includegraphics[scale=0.50]{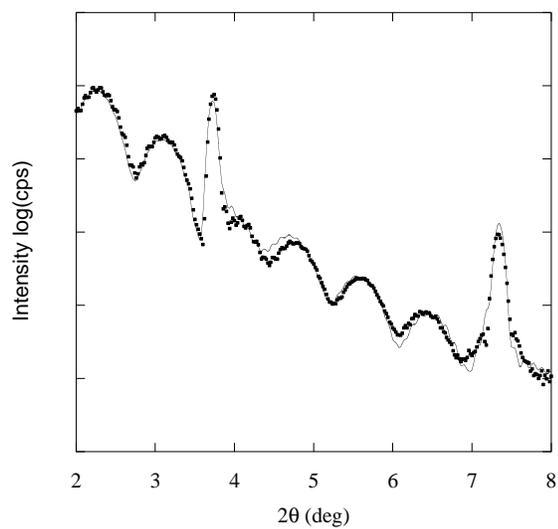}
\caption{Low-angle x-ray diffraction scan (specular component) for the Fe(3 ML)/V(13 ML) superlattice. The solid line is the fit to the measured data (filled squares).}
\label{figxray2}
\end{figure}

\begin{figure}
\includegraphics[scale=0.60]{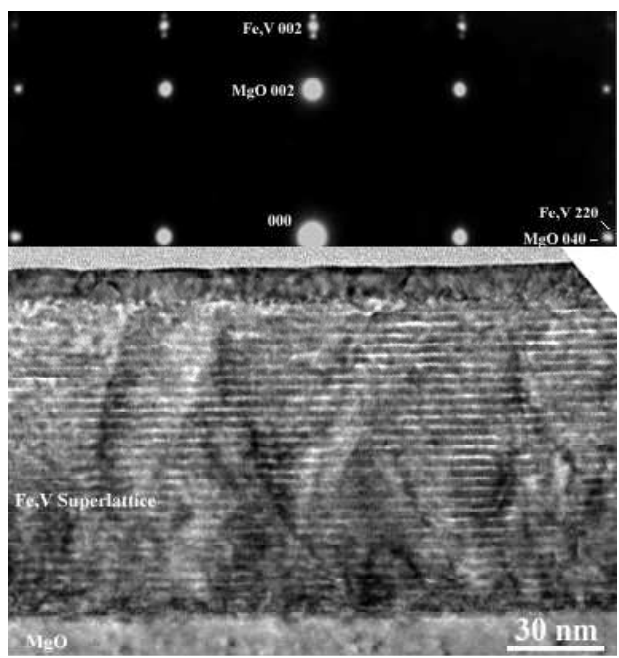}
\vspace*{0.2cm}
\caption{SAED pattern (top) and TEM micrograph (bottom) of the Fe(9 ML)/V(13 ML) superlattice.}
\label{TEM}
\end{figure}

\begin{figure}
\includegraphics[scale=0.40]{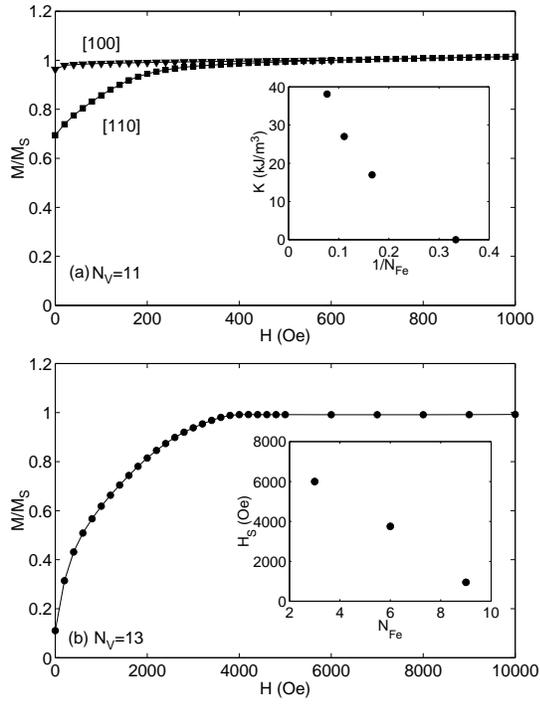}
\caption{Magnetization vs. magnetic field for (a) Fe(6 ML)/V(11 ML) and (b) Fe(6 ML)/V(13 ML); $T$=10K. For the FM coupled SL (insert (a)), the variation of the anisotropy constant $K$ with the inverse of the Fe thickness is included. In the AF case (insert (b)), the variation of the saturation field is added.}
\label{mh}
\end{figure}

A high-angle radial scan of the Fe(3 ML)/V(13 ML) superlattice is shown in Fig.~\ref{figxray1}. The peak at 62.5$^o$ is the Fe/V (002) Bragg peak, surrounded by five satellite peaks which originate from the superlattice periodicity. The peaks are sharp and well defined indicating a high structural quality of the sample. The structural coherence length ($\zeta$) in the growth direction can be estimated from the linewidth of the Bragg peak using $\zeta$=1/$\Delta q$, where $\Delta q$ is the linewidth (FWHM in \AA$^{-1}$) in the radial direction, $q=2sin \theta / \lambda$ is the scattering vector and $\theta$ is the angle of the incident and the diffracted x-rays with respect to the sample. An out-of-plane structural coherence length of about  400 \AA was obtained for the Fe(3 ML)/V(13 ML) superlattice. No other peaks than those seen in Fig.~\ref{figxray1} were detected in the range $2 \theta$=20-100$^o$ except reflections from the substrate. Furthermore, a texture scan performed on the Fe(3 ML)/V(13 ML) superlattice showed four (220) peaks separated by 90$^o$ indicating a single-crystalline superlattice.\\
The SUPREX model was used to determine the Fe/V interface roughness. The specular component of the low-angle x-ray diffraction data from the Fe(3 ML)/V(13 ML) superlattice in the range $2 \theta$=2-8$^o$ is shown in Fig.~\ref{figxray2}. The result from the fitting procedure is also shown in the figure where two distinct superlattice satellites are clearly visible. The decrease in intensity of the satellites corresponds to an average interface roughness of about two atomic monolayers ($\sim$ 3 \AA). Furthermore, the results from the fit also indicate that the Fe-on-V (Fe deposited on V) interfaces have a somewhat larger roughness than the V-on-Fe interfaces, a result which is consistent with a recent M\"ossbauer investigation of the Fe/V interfaces\cite{refmoss}. It should be pointed out that x-ray diffraction furnish structural information averaged over length scales corresponding to the coherence length of one photon. In the x-ray diffraction setup that was used, the effective in-plane coherence length of the radiation at low angles is limited by the spectral resolution $\Delta \lambda / \lambda$, to about 1000 \AA. This means that we are measuring random interface roughness as well as correlated roughness induced by the substrate. Reflection high energy electron diffraction (RHEED) patterns of the Fe and V surfaces indicate a two-dimensional layer by layer growth of both materials.\\
The TEM micrograph in Fig.~\ref{TEM} shows a cross-section of the Fe(9 ML)/V(13 ML) superlattice. The superlattice exhibits flat layers, with no significant thickness fluctuations or waviness. Superlattice satellite reflections are observed around the Fe/V (002) diffraction spot in the selected area electron diffraction (SAED) pattern shown in the figure. It is also evident that the specimen is single-crystalline and the epitaxial relationship between the superlattice and the substrate is Fe/V [001] $\parallel$ MgO [001] and Fe/V [110] $\parallel$ MgO [010], as discussed above. A detailed TEM investigation of the interface quality will be further performed.\\
As exemplified in Fig.~\ref{mh} (a) for Fe(6 ML)/V(11 ML), the SL with a V layer thickness of 11 monolayers, all
showed four-fold in-plane anisotropy, with [100] as the easy axis, except for Fe(3
 ML)/V(11 ML), which appeared isotropic in-plane. This sample saturated
at very low fields, which also excludes antiferromagnetic (AF) coupling. 
The difference in magnetocrystalline anisotropy energy ($E_a$)
between the [110] and the [100] directions $\Delta E_a=E_a[110]-E_a[100]$, proportional to the anisotropy constant $K$, increased with the thickness of the magnetic layers. $K$, however, does not exhibit the $K=K_v+2K_s/t_{Fe}$ dependence ($t_{Fe}$ is the thickness of the magnetic layers,
$K_v$ the volume and $K_s$ the 
surface coefficient of the magnetocrystalline anisotropy) which is expected if all SL are equally strained (cf. insert (a)).
The observed deviations from such a behavior are probably due to the magnetoelastic
coupling since the Fe layers are strained to accommodate a common
in-plane lattice parameter with the V layers.\cite{trem}

Fe(3, 6 and 9 ML)/V(13 ML) are magnetically isotropic in-plane, but 
rather large fields are required to reach saturation, implying that the Fe
layers are antiferromagnetically
coupled. In Fig.~\ref{mh} (b) the magnetization curve for the Fe(6 ML)/V(13 ML) is plotted. The saturation field is considerably 
larger for the isotropic V(13 ML) SL than for the V(11 ML) SL measured along the 
hard [110] direction. For a larger Fe thickness, Fe(13 ML)/V(13 ML), the 
superlattice shows a restored four-fold in-plane anisotropy and the magnetization 
curves for this sample is very similar to the corresponding curve for the  
Fe(13 ML)/V(11 ML) SL. This shows that magnetization wise,  the AF 
interlayer coupling
has become unresolvably weak for Fe layers in the thickness range 9-13 ML; as seen in the insert (b), the saturation field of the AF coupled SL drastically decreases when increasing the amount of Fe in the superlattice.

\begin{figure}
\includegraphics[scale=0.40]{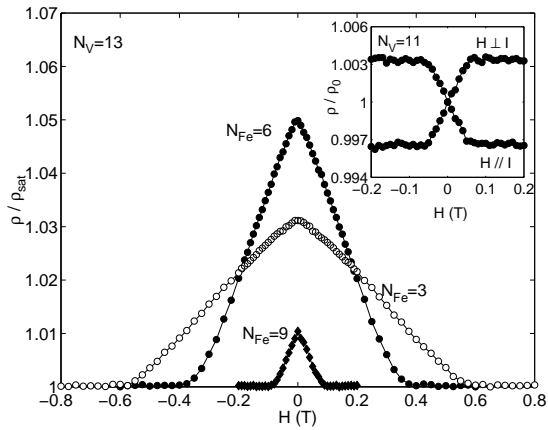}
\caption{Magnetoresistance curves for the AF coupled SL (main frame) and for one of the FM coupled SL (insert).}
\label{gmr}
\end {figure}

The magnetic field dependence of the normalized resistivity ($\rho / \rho_{sat}$ or $\rho / \rho_0$) for the Fe(3, 6 and 9 ML)/V(13 ML) SL, as well as for the Fe(13 ML)/V(11 ML) sample are shown in
Fig.~\ref{gmr}. GMR is observed for the first three samples, superposed with an
increasing anisotropic magnetoresistance (AMR) component of same order of magnitude as for the
corresponding sample in the V(11 ML) series. Fe(13 ML)/V(13 ML), on the other hand, 
displays like Fe(13 ML)/V(11 ML) only AMR features. The saturation fields derived from the GMR curves agree 
with the corresponding values derived from the magnetization measurements for the SL 
with thin Fe layers, whereas for the Fe(9 ML)/V(13 ML), a saturation field is clearly 
seen in the MR behavior but is not resolvable in the magnetization curves. The corresponding value of $\Delta \rho/\rho_0$ for the current Fe(3 ML)/V(13 ML) is
lower than a
previously reported value\cite{gbergaf} (3\% compared to 7\%), which could be
attributed to the
difference in thickness of the capping layer of Pd used, (100 {\AA}
compared to 30 {\AA} in the earlier study) as will be discussed below. This changes the amount of current going
through the superlattice, which in turn affects the MR ratio. In addition,
changing the thicknesses of the capping layer may
also change the coupling strength.\cite{brunocap} This effect
is expected to be of less importance in our case.

The SL in the V(11 ML) series show only AMR, as shown in the insert of Fig.~\ref{gmr}. The
size of $\Delta \rho/\rho_0$ is increasing with increasing Fe layer
thickness, as may be expected from the increased magnetic layer thickness and 
increasing magnetocrystalline anisotropy.

\begin{figure}
\includegraphics[scale=0.40]{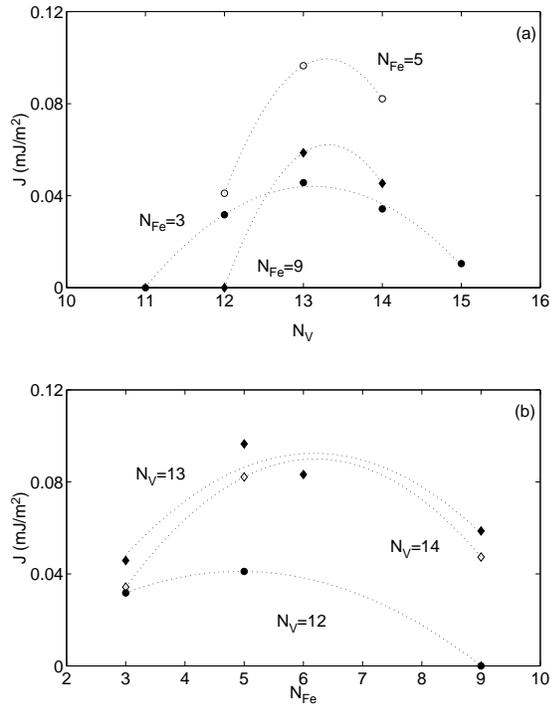}
\caption{Coupling strength of the antiferromagnetically coupled SL for (a) varying V thickness and (b) varying Fe thickness. The dotted lines are a guide to the eyes.}
\label{j}
\end{figure}

In Fig.~\ref{j} (a) the AF interlayer coupling strength is plotted vs. the number of 
V monolayers for SL with 3, 5 and 9 ML of Fe. The three series of Fe/V SL 
show a maximum  of the AF 
coupling strength at a V thickness of about 13 ML. In Fig.~\ref{j} (b), the AF coupling
strength is plotted vs. the Fe layer thickness for series of SL with V
thicknesses of 12, 13 and 14 ML. The AF coupling strength is weaker at a V thickness of 
12 ML, but all series of Fe/V SL show similar trends for 
the dependence of the coupling strength on the Fe thickness, a shallow 
maximum at about 6 ML of Fe may be estimated for all V thicknesses.

\begin{figure}
\includegraphics[scale=0.40]{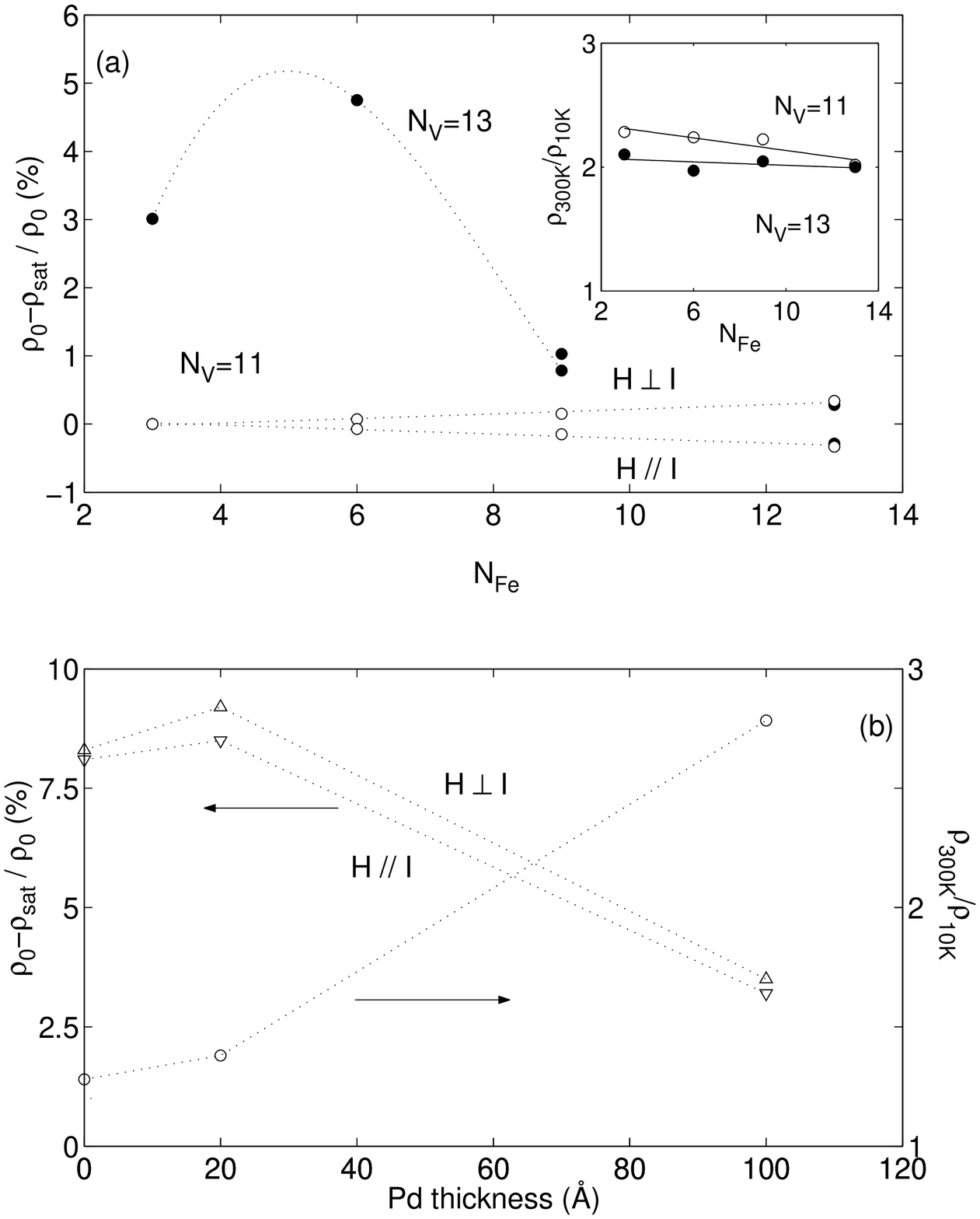}
\caption{(a) Magnitude of the GMR and AMR vs. N$_{Fe}$ for all SL. The insert shows the corresponding $\rho_{300K} / \rho_{10K}$ resistivity ratio. (b) illustrates the contribution of the Pd capping layer to the (magneto)resistivity, showing the GMR (right) and resistivity (left) ratios of Fe(5 ML)/V(13 ML) SL with respectively 0, 20, and 100 \AA of Pd. The dotted lines are a guide to the eyes.}
\label{mr}
\end {figure}

The MR values of SL in the V(11 ML) and V(13 ML) series are
displayed in Fig.~\ref{mr} (a). It is worth to note that the two Fe(13 ML) SL show an almost identical behavior only exhibiting an AMR effect. The GMR for the V(13 ML) series show a maximum of about 5$\%$ at an Fe layer thickness of about 6 ML. 

The Fe/V SL included in this study show a zero magnetic field resistivity ratio between 
300 K and 10 K [$\rho (300K)/\rho (10K)$] of about 2 (see inset of Fig.~\ref{mr}). We have observed that the resistivity 
ratio and the magnitude of the measured GMR ratio show a considerable co-variation. A 
larger resistivity ratio yields a lower GMR value for nominally similar Fe/V SL. One
obvious reason behind this behavior is as mentioned above  a difference in  thickness 
of the Pd capping layer that we always grow to protect the Fe/V SL from oxidation. Fig.~\ref{mr}(b) shows the effect of this layer on the (magneto)resistivity for an Fe(5 ML)/V(13 ML) SL. Without Pd, or for a small thickness of Pd, the resistivity ratio between room and helium temperatures amounts to $\sim$ 1.3. It increases by more than a factor of two for $\sim$ 100 \AA of Pd. At the same time the magnitude of the GMR effect drops from $\sim$ 8-9\% to $\sim$ 3\%. Because of this large variation, we consider in Fig.~\ref{mr}(a) only the SL showing similar resistivity ratios, as seen in the insert; the GMR values in this plot are thus directly comparable between each other. Of course, differences in the crystalline quality of the films and interfaces may influence the GMR magnitude.\cite{newref}

\section{CONCLUSIONS}
The interlayer exchange coupling of Fe/V (001) superlattices shows a first 
antiferromagnetic maximum at a V thickness of about 13 ML. At a V thickness of 
13 ML, the coupling has a maximum strength at an Fe layer thickness of about 6 ML. The
magnetoresistance shows GMR effects for the AF coupled SL with a maximum magnitude 
at the Fe thickness where the AF coupling is largest. It is noted that nominally
similar Fe/V SL with similar magnitude of the AF coupling can show remarkably
different GMR ratios. One simple explanation for this behavior is shunting through a 
metallic capping layer of different thickness. Measurements on a series of Fe/V
superlattices with a V thickness of 11 ML showed four fold in-plane anisotropy and only 
anisotropic magnetoresistance for all Fe layer thicknesses. An AMR of the same 
magnitude was seen superposed on the GMR effect for the series of SL with a
V thickness of 13 ML.

\begin{acknowledgments}

Financial support from the Swedish natural Science Research Council (NFR).

\end{acknowledgments}

\begin{table}
\caption{Data on the Fe($X$ ML)/V($Y$ ML) superlattices: Number of atomic monolayers and repetitions, Pd thickness, as well as nominal and measured superlattice period $\Lambda$.}
\label{table1}
\begin{tabular}{crcc}
Superlattice&Pd (\AA)&Nominal $\Lambda$ (\AA)& Measured $\Lambda$ (\AA)\\
\colrule
3/11$\times$30&100&20.95&20.80\\
3/12$\times$30&100&22.46&22.95\\
3/13$\times$30&100&23.98&24.30\\
3/14$\times$30&100&25.49&25.85\\
3/15$\times$30&100&27.00&27.65\\
5/12$\times$30&100&25.33&25.30\\
\colrule
5/13$\times$30&100&26.84&27.00\\
5/13$\times$30&20&26.84&26.50\\
5/13$\times$30&0&26.84&26.60\\
\colrule
5/14$\times$30&100&28.36&28.25\\
6/11$\times$30&100&25.25&25.20\\
6/13$\times$30&100&28.28&28.30\\
9/11$\times$30&100&29.55&28.35\\
9/12$\times$30&100&31.06&31.05\\
9/13$\times$30&100&32.58&32.80\\
9/14$\times$30&100&34.09&34.15\\
13/11$\times$30&100&35.28&36.70\\
\end{tabular}
\end{table}

\newpage
\vspace*{-2cm}

\begin{thebibliography}{0}

\bibitem{gmr}
B. Dieny, J. Magn. Magn. Mater. {\bf 136}, 335 (1994).
\bibitem{gmr2}
 P.M. Levy, Sol. State Phys. {\bf 47}, 367 (1994).
\bibitem{osc}
S. S. P. Parkin, N. More, K. P. Roche, \prl {\bf 64}, 2304 (1990).

\bibitem{iec}
P. Bruno, \prb {\bf 52}, 411 (1995); P. J. H. Bloemen, M. T. Johnson,
M. T. H. van de Vorst, R. Coehoorn, J. J. de Vries, R. Jungblut, J. aan
de Stegge, A. Reinders and W. J. M. de Jonge, \prl {\bf 72}, 764 (1994);
P. Lang, L. Nordstr\"om, K. Wildeberger, R. Zeller, P. H. Dedrichs, T.
Hoshino \prb {\bf 53}, 9092 (1996);
P. Bruno, J. Phys.: Condens. Matter {\bf 11}, 9403 (1999).

\bibitem{gbergaf}
P. Granberg, P. Isberg, E. B. Svedberg, B. Hj\"orvarsson, P. Nordblad
and R. W\"appling, J. Magn. Magn. Mater. {\bf 186}, 154 (1998).

\bibitem{refsuprex}E. E. Fullerton, I. K. Schuller, H. Vanderstraeten and Y. Bruynseraede, Phys. Rev. B {\bf 45}, 9292 (1992).

\bibitem{schwickert}
M. M. Schwickert, R. Coehorn, M. A. Tomaz, E. Mayo, D. Lederman, W. L.
O'Brien, Tao Lin and G. R. Harp, \prb {\bf 57}, 13 681 (1998).

\bibitem{refmoss}B. Kalska, P. Blomqvist, L. H\"aggstr\"om and R. W\"appling, Europhys. Lett. {\bf 53}, 395-400 (2001).

\bibitem{trem}
\'E. du Tr\'emolet de Lacheisserie, Ann. Phys. {\bf 5}, 267 (1970);
\'E. du Tr\'emolet de Lacheisserie, \prb {\bf 51}, 15925 (1995).

\bibitem{gberg15ml}
P. Granberg, P. Nordblad, P. Isberg, B. Hj\"orvarsson and R.
W\"appling, \prb {\bf 54}, 1199 (1996).

\bibitem{fevtrem}
A. Broddefalk, P. Nordblad, P. Blomqvist, P. Isberg, R. W\"appling, O. Le Bacq and O. Eriksson, {\it to appear in J. Magn. Magn. Mater.}

\bibitem{brunocap}
A. Bounouh, P. Beauvillain, P. Bruno, C. Chappert, R. M\'egy and P.
Veillet, Europhys. Lett. {\bf 33}, 315 (1996).

\bibitem{newref}
A. Moser, U. Krey, A. Paintner and B. Zellermann, J. Magn. Magn. Mater. {\bf 183}, 272 (1998).

\end{thebibliography}
\end{document}